\documentclass{article}
\usepackage{spconf,amsmath,graphicx}
\usepackage{multirow}
\usepackage{booktabs}
\usepackage{algorithm}  
\usepackage{algpseudocode}  
\usepackage{amsthm,amsmath,amssymb}
\usepackage{mathrsfs}
\usepackage{color}
\usepackage{xcolor}
\usepackage{bm}
\usepackage{bbding}

\usepackage[normalem]{ulem}

\title{Mixed Precision DNN Quantization for Overlapped \\
Speech Separation and recognition}
\name{Junhao Xu*$^1$, Jianwei Yu*$^2$\thanks{* Equal Contribution. This work is partly done when Jianwei Yu is an intern in Tencent AI lab},  Xunying Liu$^1$, Helen Meng$^1$}
\address{$^1$The Chinese University of Hong Kong; $^2$Tencent AI lab}
\begin{document}
\ninept
\maketitle
\vspace{-2ex}
\begin{abstract}
Recognition of overlapped speech has been a highly challenging task to date. State-of-the-art multi-channel speech separation system are becoming increasingly complex and expensive for practical applications. To this end, low-bit neural network quantization provides a powerful solution to dramatically reduce their model size. However, current quantization methods are based on uniform precision and fail to account for the varying performance sensitivity at different model components to quantization errors. In this paper, novel mixed precision DNN quantization methods are proposed by applying locally variable bit-widths to individual TCN components of a TF masking based multi-channel speech separation system. The optimal local precision settings are automatically learned using three techniques. The first two approaches utilize quantization sensitivity metrics based on either the mean square error (MSE) loss function curvature, or the KL-divergence measured between full precision and quantized separation models. The third approach is based on mixed precision neural architecture search. Experiments conducted on the LRS3-TED corpus simulated overlapped speech data suggest that the proposed mixed precision quantization techniques consistently outperform the uniform precision baseline speech separation systems of comparable bit-widths in terms of SI-SNR and PESQ scores as well as word error rate (WER) reductions up to 2.88\% absolute (8\% relative). 
\end{abstract}
\vspace{-1.5ex}
\section{Introduction}
\label{sec:intro}
\vspace{-0.5ex}
Despite the rapid progress of automatic speech recognition (ASR) in the past few decades, accurate recognition of overlapped speech remains a highly challenging task to date. To this end, microphone arrays and the required multi-channel signal integration technologies represented by TF masking~\cite{bahmaninezhad2019comprehensive, chen2019multi}, delay and sum~\cite{van1988beamforming, anguera2007acoustic} and minimum variance distortionless response (MVDR)~\cite{pados2001iterative, souden2009optimal} play a key role in state-of-the-art overlapped speech separation and recognition systems. With the wider application of deep learning based speech technologies, these speech separation methods have evolved and been integrated into a variety of DNN based designs based on, for example, convolutional time-domain audio separation networks (Conv-TasNets)~\cite{luo2019conv}, dual path recurrent neural networks and transformers
~\cite{luo2020dual, li2021dual}
. State-of-the-art speech separation performance require increasingly complex neural architecture designs. 
For example, the audio-only and audio-visual speech separation systems introduced in~\cite{gu2020multi} contain 9.6 and 22 million parameters in total respectively.
However, this not only lead to a large increase in their overall memory footprint and computational cost when operating on the cloud, but also creates difficulty when deployed on edge devices to enhance privacy and reduce latency.

To this end, one efficient and powerful solution is to use low-bit deep neural network (DNN) quantization techniques~\cite{KYu-2020,YQian-2019,CLeng-2018,RMa-2019}, which has drawn increasing interest in the machine learning and speech technology community in recent years. By replacing floating point weights with low precision values, the resulting quantization methods can significantly reduce the model size and inference time without modifying the model architectures. Traditional DNN quantization approaches~\cite{Courbariaux-2015, Jun-2020, fasoli21_interspeech} are predominantly based on uniform precision, where a manually defined identical bit-width is applied to all weight parameters during quantization. This fails to account for the varying performance sensitivity at different parts of the system to quantization errors. In practice, this often leads to large performance degradation against full precision models. 

In order to address the above issue, novel mixed precision DNN quantization methods are proposed in this paper to address this problem by applying locally variable bit-widths settings to individual TCN components of a TF masking based multi-channel speech separation system~\cite{yu2021audio}. These methods are becoming well supported by the recent development of mixed precision DNN acceleration hardware that allow multiple locally set precision settings to be used~\cite{ZDong-2019-2}. The resulting flexibility can provide a better trade-off between compression ratio and accuracy performance target. The optimal local precision settings are automatically learned using three techniques. The first two approaches utilize quantization sensitivity metrics based on either the mean square error (MSE) loss function curvature that can be approximated efficiently via matrix free techniques, or the KL-divergence measured between full precision and quantized separation models. The third approach is based on mixed precision neural architecture search. 

Experiments conducted on the Lip Reading Sentences based on TED videos (LRS3-TED) corpus~\cite{afouras2018deep} simulated overlapped speech data suggest that the proposed mixed precision quantization techniques consistently outperform the uniform precision baseline speech separation systems of comparable quantization bit-widths. Consistent performance improvements in terms of both SI-SNR and PESQ based speech enhancement metrics and speech recognition word error rate (WER) up to 2.88\% absolute (8\% relative) were obtained. The 8-bit KL mixed precision quantized system achieved a “lossless” quantization over the full precision 32-bit baseline while incurring no statistically significant WER increase.

The main contributions of this paper are summarized as follows. First, this paper is the first work to apply mixed precision quantization methods to speech separation tasks. In contrast, previous researches on low-bit quantization within the speech community largely focused on the back-end recognition system~\cite{fasoli21_interspeech, nguyen2020quantization} and language models~\cite{Jun-2021, XLiu-2018}. In addition, prior researches on light weight speech enhancement approaches were based on neural structural sparsity compression~\cite{tan2021towards} rather than the proposed mixed precision low-bit quantization methods. Second, the proposed 8-bit KL mixed precision quantized speech separation system achieved a “lossless” quantization over the full precision 32-bit baseline in terms of speech recognition accuracy.

The rest of this paper is organized as follows. The multi-channel speech separation system is reviewed in section 2. Uniform precision neural network quantization methods are introduced in section 3. Section 4 presents mixed precision quantization methods. Experiments and results are shown in section 5. Finally, conclusions and future work are discussed in section 6. 

\vspace{-0.5em}
\section{TF masking based Multi-channel Speech Separation }
\vspace{-1ex}
\label{avspeechsep}
This section introduces the TF masking based multi-channel speech separation framework used in this paper.

\vspace{-1.5ex}
\subsection{Audio inputs}
\vspace{-0.5ex}
As is illustrated in Figure \ref{fig:per1}, three types of audio features including the complex spectrum, the inter-microphone phase differences (IPDs) \cite{yoshioka2018recognizing} and location-guided angle feature (AF) \cite{chen2018multi,gu2019end} are adopted as the audio inputs. The complex spectrum of all the microphone array channels are first computed through short-time Fourier transform (STFT).

\noindent\textbf{IPDs features} were used to capture the relative phase difference between different microphone channels and provide additional spatial cues for TF masking based multi-channel speech separation.  These can be computed as follows:
\vspace{-0.5em}
\begin{equation}
{\rm IPD}^{(m,n)}_{t,f} = \angle({y^m_{t,f}}/{y^n_{t,f}}) 
\end{equation}
\vspace{-0.15em}
where \(y^m_{t,f}\) and \(y^n_{t,f}\) denote the STFT's TF bins of mixed speech at time $t$ and frequency bin $f$ on $m$-th and $n$-th microphone channels, respectively. The operator $\angle(\cdot)$ outputs the angle between them.
 
\noindent\textbf{Angle features} that are based on the approximated direction of arrival (DOA) were also incorporated to provide further spatial filtering constraint. In this work, the approximate DOA of a target speaker, $\theta$, is obtained by tracking the speaker’s face from a 180-degree wide-angle camera, as is shown in the left hand side of Figure \ref{fig:per1}. This allows the array steering vector corresponding to the target speaker to be expressed as follows:
\vspace{-0.5em}
\begin{align}
    {\bf{G}}(f) =  \left[e^{-j \phi_1\cos(\theta)}, e^{-j\phi_r\cos(\theta)}, ...,e^{-j\phi_R\cos(\theta)}\right]
\end{align}
\vspace{-0.15em}
where $\phi_r = 2\pi f d_{1r}/c$ and $d_{1r}$ is the distance between the  first (reference) and $r$th microphone ($d_{11} = 0$). $c$ is the sound velocity.
Based on the computed steering vector, the location-guided AF feature introduced in \cite{chen2018multi, gu2020multi} are also adopted to provide further discriminative information for the target speaker as follows:
\vspace{-0.5em}
\begin{align}
    \text{AF}(t,f) = \sum_{\{(m,n)\}} \frac{\big{\langle}\text{vec}\big{(}\frac{G_n(f)}{G_m(f)}\big{)}, \text{vec}\big{(}\frac{y^{m}_{t,f}}{y^{n}_{t,f}}\big{)}\big{\rangle}}{\big{\|}\text{vec}\big{(}\frac{G_n(f)}{G_m(f)}\big{)}\big{\|}\cdot \big{\|}\text{vec}\big{(}\frac{y^{m}_{t,f}}{y^{n}_{t,f}}\big{)}\big{\|}}
\end{align}
\vspace{-0.15em}
where $\|\cdot\|$ denotes the vector norm, $\langle\cdot,\cdot\rangle$ represents the inner product and $\{(m,n)\}$ denotes the selected microphone pairs.
$\text{vec}(\cdot)$ transforms the complex value into a 2-D vector, where the real and imaginary parts are regarded as the two vector components.

\vspace{-1ex}
\subsection{Conv-Tasnet block}
\vspace{-0.5ex}
Following previous researches on audio-visual multi-channel speech separation~\cite{luo2019conv, yu2021audio}, the temporal convolutional network (TCN) architecture, which uses a long reception field to capture more sufficient contextual information, is adopted in our separation front-ends. 
As shown in Figure 1, each TCN block is stacked by 8 Dilated 1-D ConvBlock with exponentially increased dilation factors $2^0, 2^1, .... ,2^7$.
As shown in Figure 2, the log-power spectrum (LPS) features of the reference microphone channel were initially concatenated with the IPDs and AF features before being fed into a stack of several TCN blocks to estimate the complex TF mask.

\vspace{-1em}
\begin{figure}[htb]
  \centering
  \includegraphics[width=5.5cm]{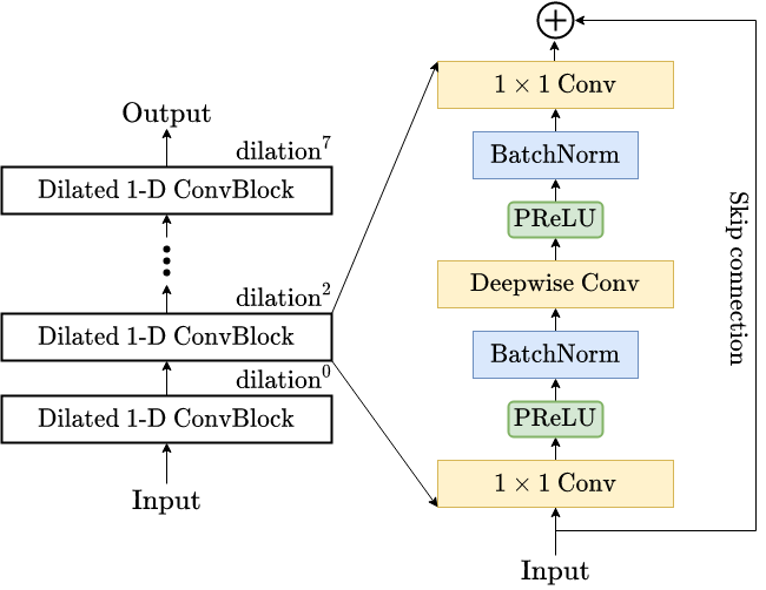} 
  \vspace{-0.5em}
  \label{fig:sub-first}
\caption{\emph{An example temporal convolutional network (TCN). Each dilated 1-D ConvBlock consists of a 1$\times$1 convolutional layer, a depth-wise separable convolution layer (D--Conv)~\cite{chollet2017xception}, with PReLU~\cite{he2015delving} activation and normalization added between convolution layers, and skip connections added between dilated 1-D ConvBlocks.}}
\label{fig:fig}
\end{figure}
\vspace{-2em}

\vspace{-1ex}
\subsection{TF masking based speech separation}
\vspace{-0.5ex}
Previous researches suggest that the complex ratio masks (CRMs) outperform both the binary masks (BMs) and real-value ratio masks (RMs) on speech separation~\cite{xu2019joint, williamson2015complex} and enhancement~\cite{hu2020dccrn} tasks. 
For this reason, the complex ideal ratio mask (cIRM) $m_{t,f}$ of the target speech is estimated in the separation module. The estimated target speech complex spectrum is obtained as:
\begin{equation}
x_{t,f} = m_{t,f} y_{t,f}^{r} \label{2.1}
\end{equation}
where $m_{t,f} \in \mathbb{C}$ is the cIRM of the target speaker, $y_{t,f}^{r}$ is the reference channel complex spectrum of mixed speech (without loss of generality, the first channel is selected as the reference channel throughout this paper). Given the estimated complex spectrum, the time-domain separated speech can be computed by the inverse short-time Fourier transform (iSTFT) and  the SI-SNR cost function is used to optimize the separation neural networks.

\begin{figure*}[h]
    \begin{center}
    \includegraphics[width=15cm]{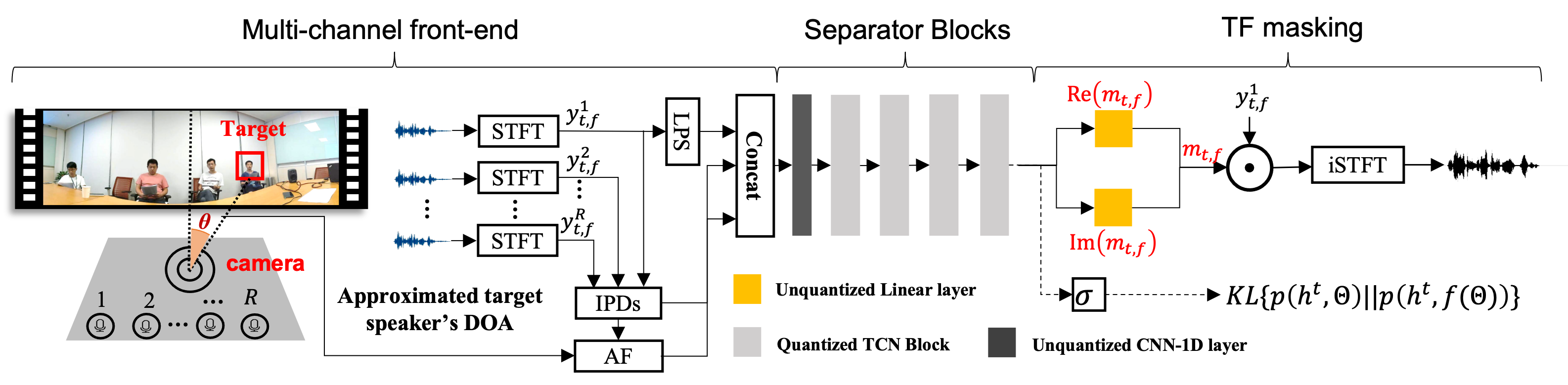}
     \vspace{-1,5em}
    \caption{\emph{Illustration of the proposed quantized audio-visual multi-channel speech separation networks, where $y^r_{t,f}$ is the complex spectrum of each channel. For the channel integration approach TF masking, $m_{t,f}$ denotes the complex mask of the target speaker and ${\rm Re}{(m_{t,f})}$ and ${\rm Im}(m_{t,f})$ are the real and imaginary part. The quantized TCN blocks are marked light grey in the figure.}}
    \label{fig:per1}
        \vspace{-2.5em}
    \end{center}
\end{figure*}

\vspace{-1.5ex}
\section{Neural Network Quantization}
\vspace{-0.5ex}
\label{sec:nnquantization}
For a standard $n$-bit quantization problem of neural networks, we consider a full precision weight parameter $\Theta$ and find its closest discrete approximation from the following quantization table $Q \in \{0, \pm1,\pm2, \dots, \pm (2^{n-1}-1)\}$ as
\begin{equation}
 f(\Theta) = \arg\min \limits_{Q} |\Theta - Q|
\end{equation}
while one bit is reserved to denote the sign. With further simplification, low bit quantization, for example, binarization $\{1, -1\}$~\cite{Rastegari-2016, CLeng-2018} and ternary $\{-1, 0, 1\}$~\cite{LiF-2016}, can be produced.

When applying quantization to all weight matrices in the model, we can use a more general format in equation (5) to represent the quantization for each parameter. Let $\Theta^{(l)}_{i}$ be the $i^{th}$ parameter within any of the $l^{th}$ weight cluster, for example, all weight parameters of the same Conv-Tasnet layer,

\vspace{-1em}
 \begin{equation}
  \begin{split}
  \label{eq:quanMap}
  f(\Theta^{(l)}_i) = \arg\min \limits_{Q_i^{(l)}} |\Theta^{(l)}_i - Q_i^{(l)}|
\vspace{-1em} 
\end{split}
\end{equation}
\vspace{-1em}

The locally shared $l^{th}$ quantization table is given by

\begin{equation}
\label{equ:31}
  {Q}_i^{(l)}\in\{0, \pm\alpha^{(l)},\dots, \pm\alpha^{(l)} (2^{n_{l}-1}-1)\}
\end{equation}
where $\alpha^{(l)}$ is a full precision scaling factor used to adjust the dynamic range of all the unquantized weights in the cluster. It is shared locally among weight parameters clusters. A special case, when the local quantization table in equation (7) is shared across all the layers, leads to the traditional uniform precision quantization approach. The only remaining factor affecting the system performance is the bit length $n_{l}$ which is also globally set to be $1, 2, 4, 8, 16$ etc.

\begin{table*}[h]
  \caption{\emph{Performance of the baseline full precision, uniform precision quantized and mixed preciison quantized TF-masking based speech separation systems with local precision settings automatically learned from HES/KL/NAS introduced in section 4 on LRS3-TED corpus. WERs measured both on average and across subsets of test data with varying between speaker angles from 0-15 to 90-180 degrees. ${\star}$, ${\dag}$  and ${\ddag}$ denote no statistically significant WER difference obtained over baseline systems (sys.1,3,4). Evaluation time in seconds per hour of speech.} }
  \label{tab:swbd-trans}
  \centering
\resizebox{175mm}{20mm}{
  \begin{tabular}{c|c|c|
  c|c|c|cccc|ccccc|c|c} 
      \toprule
      \multirow{2}{*}{\textbf{Sys}}&  \textbf{quant.}&\textbf{param.} & \textbf{prec.} & \textbf{quant.}&\multirow{2}{*}{\textbf{\#bit}} & 
      \multicolumn{4}{c|}{\textbf{SI-SNR(PESQ)}}& \multicolumn{5}{c|}{\textbf{WER(\%)}} &
  \textbf{model} & \textbf{eval. time}  \\  
      &\textbf{prec.}& \textbf{estim.} & \textbf{set} &\textbf{method}& & 0-15 & 15-45 & 45-90 & 90-180 & 0-15 & 15-45 & 45-90 & 90-180 & avg. & \textbf{size(MB)}  & \textbf{(sec./hour)}  \\ \midrule
1 & \multicolumn{4}{c|}{baseline} & 32 & 7.13(2.65) & 10.22(3.01) & 10.94(3.10) & 10.83(3.10) & 40.55 & 26.37 & 23.85 & 23.82 & 28.65 & 35.2 & 67.79 \\ \cline{1-2}\cline{3-17}
2 &  &  & \multirow{4}{*}{-} & & 2 & 5.22(2.38) & 7.89(2.64) & 8.67(2.71) & 8.57(2.70) & 49.95 & 37.17 & 33.25 & 33.71 & 38.52 & 4.6 & 37.22 \\ \cline{1-1}
3 & \emph{uniform} &   & & \emph{manual} & 4 & 6.43(2.52) & 9.05(2.80) & 9.86(2.88) & 9.62(2.86) & 44.88 & 32.16 & 29.49 & 29.86 & 34.10 & 6.6 & 36.54 \\ \cline{1-1}
4 & \emph{prec.} &  & & \emph{define} & 8  & 7.10(2.62) & 9.94(2.94) & 10.65(3.02) & 10.46(3.01) & 41.16 & 28.09 & 25.16 & 25.58 & 30.00$\star$ & 10.7 & 36.98 \\ \cline{1-1}
5 & &   & & & 16  & 7.11(2.64) & 10.22(3.01) & 10.91(3.09) & 10.81(3.09) & 40.81 & 26.00 & 23.95 & 23.87 & 28.66${\star}$ & 18.9 & 54.74 \\ \cline{1-2}\cline{4-17}
6 & & Post-training & \multirow{6}{*}{$\{2,4,8,16\}$} & \multirow{2}{*}{Hes} & 4 & 6.66(2.56) & 9.60(2.89) & 10.29(2.96) & 10.09(2.95) & 42.91 & 28.90 & 26.47 & 27.17 & 31.36${\ddag}$ & 6.7 & 44.28  \\ \cline{1-1}
7 &  & Offline & & & 8 & 7.21(2.65) & 10.15(3.00) & 10.85(3.08) & 10.76(3.09) & 40.31 & 26.50 & 24.71 & 24.23 & 28.94${\star}$${\ddag}$ & 10.8 & 48.99  \\ \cline{1-1}\cline{5-17}
8 & \emph{mixed} & Quant~\cite{fasoli21_interspeech} & & \multirow{2}{*}{KL} & 4 & 6.89(2.59) & 9.61(2.89) & 10.35(2.97) & 10.17(2.96) &  42.01 & 29.49 & 26.42 & 26.95 & 31.22${\ddag}$ & 6.7 & 43.61  \\ \cline{1-1}
9 &  \emph{prec.} & &  & & 8 & 7.20(2.65) & 10.20(3.00) & 10.87(3.08) & 10.75(3.08) & 40.28 & 26.62 & 23.97 & 23.17 & 28.51 & 10.8 & 47.85  \\ \cline{1-1}\cline{5-17}
10 &  & & & \multirow{2}{*}{NAS} & 4 & 6.54(2.54) & 9.22(2.83) & 9.95(2.92) & 9.86(2.90) & 44.46 & 31.00 & 27.89 & 28.26 & 32.90${\dag}$ & 6.7 & 44.47 \\ \cline{1-1}
11 & & & & & 8 & 7.08(2.61) & 10.13(2.99) & 10.82(3.07) & 10.73(3.08) & 41.11 & 26.63 & 24.29 & 24.52 & 29.14${\star}$${\ddag}$ & 10.8 & 49.10
\\ 
\bottomrule
      \end{tabular}
      }
          \vspace{-1.2em}
\end{table*}

\vspace{-1.5ex}
\section{Mixed Precision Quantization}
\vspace{-0.5ex}
\label{sec:mixedquant} 
This section presents three approaches to automatically learn the optimal local precision settings for the quantization of our TF-masking based multi-channel speech separation system. 

\vspace{-1.5ex}
\subsection{KL Divergence Based Mixed Precision Quantization}
In order to minimize the distance between the distribution embodied by the original full precision system and that of the quantized model, Kullback-Leibler (KL) divergence between full precision and quantized NNs is used to measure the resulting performance sensitivity. 
Taking a $L$-layer NN for example, for any quantization $f(\cdot)$ being applied to the full precision parameters ${\bf \Theta}$, the KL divergence based quantization sensitivity measure is computed over the input spectrum of $T$ frame length as,
\vspace{-1em}
\begin{align}
  \Omega^{\rm{KL}} &= \sum_{i=1}^L\Omega^{\rm{KL}}_i = \sum_{i=1}^L D_{\rm{KL}}(P(\bm{\Theta_i})||P(f_{n_i}(\bm{\Theta_i})))\\
  & =\sum_{i=1}^{L} \sum_{t=1}^{T}P(\sigma(\bm{h}^{t}), {\bf \Theta}_i)\ln{\frac{P(\sigma(\bm{h}^{t}), {\bf \Theta }_i)}{P(\sigma(\bm{h}^{t}), f_{n_i}({\bf \Theta}_i)}}\notag
 \end{align}
 \vspace{-0.5em}
 
\noindent where ${\bf \Theta}_i$ denote the full precision parameters of the $i^{th}$ layer, and $f_{n_i}({\bf \Theta}_i)$ is $n_{i}$-bit quantized parameters given a particular local precision bit width $n_i$ and ${\bm h}^{t}$ is the TCN separator output vector computed at frame $t$. When computing the KL metric in Eqn. (8), ${\bm h}^{t}$ is fed into a Sigmoid gate (Figure 2, middle right) first to produce normalised, probability like outputs between 0 and 1.
Given a target model size constraint (e.g. average 4-bit precision), the KL metric for each precision setting of each layer is computed and minimized to select the optimal local bit-width while satisfying the constraint\footnote{A minimum 2-bit precision is also enforced for all layers during this layer by layer quantization precision optimization to filter out invalid settings.}.

 \vspace{-1.5ex}
\subsection{Curvature Based Mixed Precision Quantization}
The second approach minimizes the performance sensitivity to quantization by examining the local training data SI-SNR separation error loss function curvature. Under mild assumptions such that the parameters of a DNN is twice differentiable and while converging to a local optimum, it is shown in ~\cite{ZDong-2019-2, ZDong-2019} that the separation performance sensitivity to quantization, when using a given precision setting, can be expressed as the squared quantization error further weighted by the parameter Hessian matrix trace. For any quantization $f(\cdot)$ being applied to the parameters $\boldsymbol{\Theta}$ of the L-layer Conv-Tasnet separation model, the total performance sensitivity is given by the  sum of Hessian trace weighted squared quantization error, to be minimized under a target model size constraint.
\vspace{-0.5em}
\begin{equation}
  \Omega^{\rm{Hes}} = \sum_{i=1}^L\Omega^{\rm{Hes}}_i = \sum_{i=1}^L Tr(\bm{H}_i)\cdot||f(\bm{\Theta}_i)-\bm{\Theta}_i||_2^2
\end{equation}
\vspace{-0.5em}

\noindent
An efficient stochastic linear algebra approach based on the Huchinson’s Algorithm~\cite{Avron-2011} is used to approximate the Hessian trace, 
\vspace{-0.5em}
\begin{equation}
  Tr(\bm{H})\approx \frac{1}{m}\sum_{i=1}^m \bm{z}_i^\top\bm{H}\bm{z}_i
\end{equation} 
\vspace{-0.5em}

\noindent where the expensive matrix multiplication between $\boldsymbol{H}$ and $\bm{z}_i$ can be avoided, and efficiently computed using Hessian-free approaches~\cite{ZDong-2019}. $\bm{z}_i$ is a random vector sampled from a Gaussian Distribution $\mathcal{N}(\bf{0}, \bf{1})$.

\vspace{-1.5ex}
\subsection{Architecture Search Based Mixed Precision Quantization}
The third solution to automatically learn the optimal local quantization precision settings is to use mixed precision based neural architecture search (NAS)~\cite{Elsken-2019, Hu-2020-CVPR} approaches. 
The super-network is constructed by first separately training the speech separation system using uniform precision, e.g. 2-bit, 4-bit, 8-bit and 16-bit, before connecting these uniform precision quantized models at each layer.

In order to avoid the trivial selection of the longest, most generous quantization bit width, these precision selection weights learning can be further constrained by a model complexity penalty term with respect to the number of bits retained after quantization, in order to obtain a target average quantization precision, for example, 4-bit,
\begin{equation}
    \Omega^{\rm{NAS}}=\mathcal{L}_{SI-SNR}(\boldsymbol{\Theta}) + \beta\sum_{(n, l)} a_n^l\cdot\sqrt{n} 
       \vspace{-0.5em}
\end{equation} 
\noindent where $\mathcal{L}_{SI-SNR}(\boldsymbol{\Theta})$ is the scale-invariant signal to noise ratio (\emph{SI-SNR}) objective function. $a^l_n$ denotes the architecture weights using $n$-bit quantization for the $l$-th cluster of weight parameters. $\beta$ is a scaling factor empirically set as $0.5$ in all experiments of this paper.

\section{Experiments}
\vspace{-0.5ex}
\noindent{\bf LRS3 Corpus and overlapped speech simulation:} We adopt the Lip Reading Sentences based on TED videos (LRS3-TED), which contain both the talking faces and subtitles of what is said in this paper. The original LRS3-TED corpus is divided into three subsets: \emph{Pre-train}, \emph{Train-val} and \emph{Test} set. We keep the \emph{Train-val} subset and randomly select the same portion of each speaker as the \emph{Pre-train} subset to make up for a $100h$ corpus in total for training. Details of the simulation process is similar to~\cite{yu2021audio}. A 15-channel symmetric linear array with noneven inter-channel spacing is used in the simulation process. Reverberation is also added in the simulated data by convolving the single channel signals with the Room Impulse Responses (RIRs) generated by the image-source method. The room size is randomly selected ranging from $4\times4\times2.5\ m^3$ to $10\times8\times6\ m^3$ (length$\times$width$\times$height) and the reverberation time T60 is sampled from a range of 0.05 to 0.7s. The average overlapping ratio of the simulated utterances is around 85\% and SIR is around 0dB. The simulated data is divided into three subsets for training (141h), validation (2h) and evaluation (0.85h). 

\noindent{\bf Implementation details}: 
Details of the IPD, AF features and hyper-parameter settings of the LF-MMI CLDNN based audio-visual recognition back-end can be found in \cite{yu2021audio, shao2020pychain}. Note that the recognition back-end is trained on clean speech data.
For each TCN block of the separation front-end, the number of channels in the 1x1 Conv-layer is set to 256 for every Dilated 1-D ConvBlock. For all the D-Conv layers, the kernel size is set to 3 with 512 channels. The implementation used to evaluate the mixed precision quantization methods of this paper is based on the existing low-bit quantized precisions that are already natively supported by the NVidia Tesla V100 GPU. These include the use of the Boolean and masking operators to implement 1-bit quantization, and the INT8 data type used to implement 2, 4 and 8-bit quantization. In case of 2-bit and 4-bit quantization, extra padded bits of zero were also included. 
Statistical significance test was conducted at level $\alpha= 0.05$ based on matched pairs sentence segment word error (MAPSSWE) for recognition performance analysis.

\noindent{\bf Experiment results:}
Table 1 presents the SI-SNR, PESQ performance and word error rates (WERs) of the baseline full precision, uniform and mixed precision quantization TF-masking based multi-channel speech separation systems on the LRS3-TED corpus simulated overlapped speech data. 
There are several trends can be found. First, given the same quantization precision, for example, 4-bit, all the mixed precision quantized models proposed in Section 4, curvature based HES (sys.6), KL (sys.8) and mixed precision NAS (sys.10) outperform the 4-bit uniform quantized model (sys.3). The 4-bit and 8-bit HES (sys.6, 7), and KL (sys.8, 9) quantized systems consistently outperform the uniform precision baseline of comparable bit-widths in terms of SI-SNR and PESQ scores, as well as WER reductions up to 2.88\% absolute (sys.8 vs. sys.3, 8\% relative). Second, among all the mixed precision quantization methods, the best WER performance for 8-bit quantization is also obtained using KL (sys.9), producing a compression ratio of 3.3 over the baseline full precision model (sys.1) and no WER increase, while uniform precision quantization requires 16-bit (sys.5) to give a similar WER.

The local precision settings of the 4-bit KL mixed precision quantized system (sys.4, Table 1) is shown in Figure 3, where the first and last two TCN layers generally require longer precision than those of the intermediate layers.

\begin{figure}[h]
    \begin{center}
    \includegraphics[width=8.6cm]{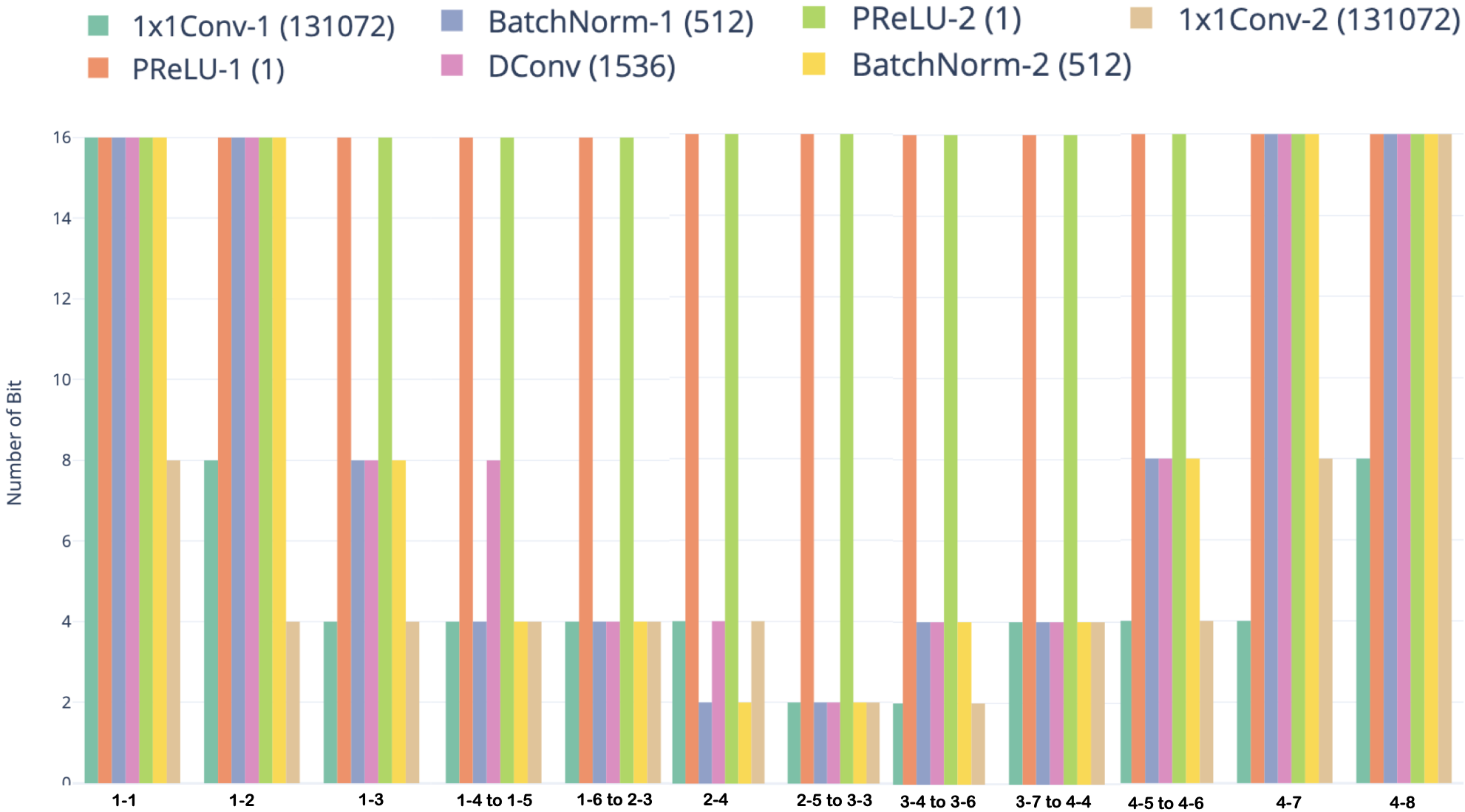}
     \vspace{-1,5em}
    \caption{\emph{Local \#bits used in avg. 4-bit KL mixed precision quantized separation TCN blocks (Figure 2, centre right in black, also as sys.8 in Table 1). TCN layer indexing $m$-$n$ denotes $m^{th}$ TCN block's $n^{th}$ dilated 1-D ConvBlock, whose 7 sublayers in Figure 1 shown in different colours together with number of parameters in brackets. "$m$-$n$ to $p$-$q$" denote consecutively positioned layers using same \#bits.}}
    \label{fig:per}
        \vspace{-3em}
    \end{center}
\end{figure}

\vspace{-2ex}
\section{Conclusions} 
\vspace{-2ex}
This paper presents novel mixed precision quantization methods for TF-masking based overlapped speech separation systems. Local precision settings are automatically learned to provide better trade-off between speech separation model compression ratio and performance loss. Experiments conducted on the LRS3-TED corpus suggest mixed precision quantization consistently outperform uniform precision quantization using comparable bit-widths. Future researches focus on improving hardware implementation and integration with back-end speech recognition systems.

\vspace{-2ex}
\section{Acknowledgement}
\vspace{-2ex}
This research is supported by Hong Kong RGC GRF grant No. 14200218, 14200220, 14200021, TRS T45-407/19N, and Innovation \& Technology Fund grant No. ITS/254/19.


\bibliographystyle{IEEEbib}
\bibliography{refs}
\end{document}